\documentclass[prd,onecolumn,showkeys,
preprint,
nofootinbib,
 amsmath,amssymb,
 aps,
]{revtex4-2}
\usepackage{yfonts}
\usepackage{graphicx}
\usepackage{dcolumn}
\usepackage{bm}
\usepackage{nicefrac}
\usepackage{xcolor}
\definecolor{darkblue}{RGB}{0,0,196}
\definecolor{darkgreen}{RGB}{0,120,0}
\usepackage[colorlinks=true,linktocpage=true,linkcolor=darkblue,citecolor=red,urlcolor=darkblue]{hyperref}
\usepackage{caption}
\usepackage[caption=false,font=footnotesize,captionskip=-2mm,farskip=-1mm]{subfig}

\usepackage{float}

\newcommand{\bea}{\begin{eqnarray}}
\newcommand{\eea}{\end{eqnarray}}
\newcommand{\bel}[1]{\begin{eqnarray}\label{#1}}
\newcommand{\eel}{\end{eqnarray}}

\def\LB{\left(}
\def\RB{\right)}
\def\LSB{\left[}
\def\RSB{\right]}

\newcommand{\nn}{\nonumber}

\newcommand{\EQ}[1]{Eq.~(\ref{#1})}
\newcommand{\EQn}[1]{(\ref{#1})}

\newcommand{\EQS}[1]{Eqs.~(\ref{#1})}

\newcommand{\EQSTWO}[2]{Eqs.~(\ref{#1})~and~(\ref{#2})}

\newcommand{\EQSM}[2]{Eqs.~(\ref{#1})--(\ref{#2})}

\newcommand{\CIT}[1]{Ref.~\citep{#1}} 
\newcommand{\CITn}[1]{\citep{#1}} 

\newcommand{\p}{\partial}
\newcommand{\dd}{\mathrm{d}}

\newcommand{\Ckv}{{\boldsymbol C}_k} 
\newcommand{\Cov}{{\boldsymbol C}_\omega}

\newcommand{\Vv}{{\boldsymbol V}}

\newcommand{\tv}{{\boldsymbol t}}

\newcommand{\f}[2]{\frac{#1}{#2}}

\def\spin{\,\textgoth{s:}}

\begin{document}


\title{Boost-invariant spin hydrodynamics with spin feedback effects 
}

\author{Zbigniew Drogosz}
\email{zbigniew.drogosz@alumni.uj.edu.pl}
\affiliation{Institute of Theoretical Physics, Jagiellonian University, PL-30-348 Krak\'ow, Poland}

\author{Wojciech Florkowski}
\email{wojciech.florkowski@uj.edu.pl}
\affiliation{Institute of Theoretical Physics, Jagiellonian University, PL-30-348 Krak\'ow, Poland}

\author{Natalia Łygan}
\email{natalia.lygan@student.uj.edu.pl}
\affiliation{Institute of Theoretical Physics, Jagiellonian University, PL-30-348 Krak\'ow, Poland}

\author{Radoslaw Ryblewski}
\email{radoslaw.ryblewski@ifj.edu.pl}
\affiliation{Institute of Nuclear Physics, Polish Academy of Sciences, PL-31-342 Krak\'ow, Poland}

\date{\today}
\bigskip
\begin{abstract}
A recently formulated extension of perfect spin hydrodynamics, which includes second-order corrections in the spin polarization tensor to the energy-momentum tensor and baryon current, is studied in the case of a one-dimensional boost-invariant expansion. The presence of second-order corrections introduces feedback from spin dynamics on the hydrodynamic background, constraining possible spin polarization configurations. However, as long as the magnitude of the spin polarization tensor remains small (below unity in natural units), the permitted spin dynamics differs very little from that found in the case without the second-order corrections. 
\end{abstract}


\maketitle
\newpage
   
\date{\today}
	
\maketitle
%
%

\section{Introduction}
\label{sec:introduction}
Experimental results showing a non-zero spin polarization of $\Lambda$ hyperons and vector mesons produced in high-energy heavy-ion collisions~\cite{STAR:2017ckg, Adam:2018ivw, Niida:2018hfw} triggered the development of spin hydrodynamics—a theoretical framework that incorporates spin dynamics into a well established theory of relativistic hydrodynamics~\cite{Florkowski:2017olj,Romatschke:2017ejr}. 
Currently, theoretical studies of spin polarization in heavy-ion collisions follow the following directions: {\bf i)} Gradients of the standard hydrodynamic variables are used on the freeze-out hypersurface only~\CITn{Becattini:2009wh,  Becattini:2011zz, Becattini:2013fla, Becattini:2013vja, Karpenko:2016jyx, Becattini:2016gvu, Becattini:2017gcx, Becattini:2021iol, Palermo:2024tza}. In this approach, primary contributions arise from the thermal vorticity $\varpi_{\mu \nu} = -\frac{1}{2} (\p_\mu \beta_\nu-\p_\nu \beta_\mu)$ and thermal shear $\xi_{\mu \nu} = -\frac{1}{2} (\p_\mu \beta_\nu+\p_\nu \beta_\mu)$, where $\beta_\mu$ is the ratio of the fluid flow vector $U_\mu$ to the local temperature $T$. {\bf ii)} The hydrodynamic equations are derived from moments of kinetic equations that incorporate spin degrees of freedom~\CITn{Florkowski:2017ruc, Florkowski:2017dyn, Florkowski:2018ahw, Florkowski:2018fap, Kumar:2018iud, Kumar:2018lok, Bhadury:2020puc, Bhadury:2020cop, Bhadury:2022ulr, Singh:2022ltu, Weickgenannt:2019dks, Weickgenannt:2021cuo, Weickgenannt:2020aaf, Wagner:2022amr, Weickgenannt:2022zxs, Weickgenannt:2023nge, Wagner:2024fhf, Hu:2021pwh, Li:2020eon, Shi:2020htn,  Singh:2022ltu}. {\bf iii)}~Spin hydrodynamics is formulated by making the reference to mathematically allowed forms for the energy-momentum and spin tensors, with conservation laws and the entropy increase law being  enforced~\CITn{Hattori:2019lfp, Fukushima:2020ucl, Daher:2022xon, Daher:2022wzf, Sarwar:2022yzs, Wang:2021ngp, Biswas:2022bht, Biswas:2023qsw, Xie:2023gbo, Daher:2024ixz, Xie:2023gbo, Ren:2024pur, Daher:2024bah, Gallegos:2021bzp, Hongo:2021ona, Kumar:2023ojl, She:2021lhe}. {\bf iv)} A spin-extended Lagrangian formalism is employed~\CITn{Montenegro:2017rbu,Montenegro:2020paq,Goncalves:2021ziy}.

In the framework of perfect spin hydrodynamics, as summarized and reviewed in Ref.~\cite{Florkowski:2018fap}, spin dynamics follows solely from the conservation of energy, linear momentum, baryon number, and the spin part of the total angular momentum. Currently, this type of approach is considered to be rooted in kinetic theory which accounts only for local collisions~\CITn{Florkowski:2017ruc, Florkowski:2017dyn, Florkowski:2018ahw, Florkowski:2018fap, Bhadury:2020puc, Bhadury:2020cop, Bhadury:2022ulr}, with locality here referring to the dominance of $s$-wave scattering. A significant extension to non-local collisions has been introduced in Refs.~\CITn{Weickgenannt:2019dks, Weickgenannt:2021cuo, Weickgenannt:2020aaf, Wagner:2022amr, Weickgenannt:2022zxs, Weickgenannt:2023nge, Wagner:2024fhf}. Non-local collisions enable exchange between the orbital and spin parts of the total angular momentum. Such processes are also responsible for bringing the spin polarization tensor $\omega_{\mu\nu}$ towards the thermal vorticity $\varpi_{\mu \nu}$ as the system approaches global equilibrium. 

In two recent papers \CITn{Florkowski:2024bfw,Drogosz:2024gzv}, the formalism of perfect spin hydrodynamics developed in Ref.~\cite{Florkowski:2018fap} was extended to incorporate dissipative phenomena. At the same time, its connection with the works following the track {\bf iii)} has been clarified. It turns out that consistency across different spin hydrodynamics formulations can be achieved if the expressions for the baryon current and the energy-momentum tensor include at least second-order terms in the spin polarization tensor $\omega$. Notably, in natural units, the spin polarization tensor $\omega_{\mu\nu}$ is dimensionless—it can be defined as the ratio of the spin chemical potential $\Omega_{\mu\nu}$ to the temperature $T$, $\omega_{\mu\nu} = \Omega_{\mu\nu}/T$. Hence, the expansion in the components of $\omega_{\mu\nu}$ is well defined. 

Since the inclusion of such second-order terms is a new aspect of spin hydrodynamics, it is interesting and important to examine their impact on the time evolution of spin-polarized systems. To study this matter, in this work we consider a boost-invariant one-dimensional expansion and neglect dissipative terms. In this way, we can explore the effects of the second-order corrections in the case of the simplest possible geometry of expansion. Given this simplified geometry, we do not make any attempts to describe experimental data. We rather concentrate on the analysis of possible effects resulting from the second-order terms. 

Our present approach can be regarded as an extension of the analysis presented in~\CIT{Florkowski:2019qdp}. Throughout the discussion, we present numerous comparisons with that work. Our main conclusion is that the addition of second-order terms strongly constrains the allowed forms of the spin tensor. On the other hand, we find small numerical effects on the evolution of the allowed configurations (at least if the values of the spin polarization tensor remain small, which is a necessary condition to justify expansion in $\omega$).

\smallskip
\noindent
{\it Notation and conventions.} We use the mostly-minuses convention for the metric tensor, $g_{\mu\nu} = \textrm{diag}(+1,-1,-1,-1)$, and the Levi-Civita tensor with $\epsilon^{0123}=+1$. Throughout the text, we use natural units, $\hbar = c = k_B = 1$.

\section{boost-invariant spin polarization tensor}

We explore herein the simplest, boost-invariant expansion geometry known as the Bjorken expansion \cite{Bjorken:1982qr}. The boost invariance imposes strong conditions on the forms of the hydrodynamic variables. Most of such conditions for spin hydrodynamics were discussed before in~\CIT{Florkowski:2019qdp}. Below we recall and generalize this analysis. 

\subsection{Spin polarization tensor}

The basic object in spin hydrodynamics is the spin polarization tensor $\omega_{\mu \nu}$, which (similarly to the Faraday tensor in classical electrodynamics) is a rank-2 antisymmetric tensor, $\omega_{\mu\nu}=-\omega_{\nu\mu}$. Following the method used in relativistic magnetohydrodynamics, we decompose the spin polarization tensor into an ``electric'' and a ``magnetic'' part~\footnote{For comparisons with magnetohydrodynamics, see, for example, \CIT{PhysRevE.51.4901}.}
\bea
\omega_{\mu \nu}=k_{\mu} U_{\nu} -k_{\nu} U_{\mu}+t_{\mu\nu} =
k_{\mu} U_{\nu} -k_{\nu} U_{\mu}+\epsilon_{\mu \nu \alpha \beta} U^{\alpha} \omega^{\beta}.
\label{eq:o1}
\eea
Here, $U$ is the hydrodynamic flow, while the four-vectors $k$ and $\omega$ (the electriclike and magneticlike components, respectively) satisfy the orthogonality conditions
\bea
k_{\mu} U^{\mu} &=& 0, \qquad \omega_{\mu} U^{\mu} = 0.
\eea
Equation~\EQn{eq:o1} implicitly defines the tensor $t_{\mu\nu} = \epsilon_{\mu \nu \alpha \beta} U^{\alpha} \omega^{\beta}$.

For boost-invariant and transversely homogeneous systems, it is customary to adopt the following orthonormal basis:
\begin{align}\begin{split}\label{eq:4_vectors}
U ^{\alpha} &= \frac{1}{\tau}(t,0,0,z)= \LB \cosh\eta,0,0, \sinh\eta \RB, \\
X ^{\alpha} &= (0,1,0,0), \\
Y ^{\alpha} &= (0,0,1,0), \\
Z ^{\alpha} &= \frac{1}{\tau}(z,0,0,t)= \LB \sinh\eta,0,0,\cosh\eta \RB,
\end{split}\end{align}
where $\tau=\sqrt{t^2 - z^2}$ is the longitudinal proper time, and $\eta= \f{1}{2} \hbox{ln} \left( \f{t+z}{t-z} \right)$ is the spacetime rapidity. It is straightforward to verify that the four-vectors \EQn{eq:4_vectors} have a boost-invariant form, i.e., they all satisfy the condition  $V'^{\mu}(x') = L^{\mu}_{\ \nu} V^{\nu}(x)= V^{\mu}(x')$, where $L^{\mu}_{\ \nu}$ represents a Lorentz boost along the $z$ axis, $V(x)$ is one of the four-vectors appearing in \EQ{eq:4_vectors}, and $x' = L \,x$.~\footnote{The boost invariance of vector fields is discussed in more detail in~\CIT{Florkowski:2010zz}.}

Since the four-vectors $k$ and $\omega$ are orthogonal to the flow vector $U$, we can decompose them as follows:
\bea
k^{\mu} = C_{kx}X^{\mu} + C_{ky}Y^{\mu} + C_{kz}Z^{\mu}, \quad
\omega^{\mu} = C_{\omega z}X^{\mu} + C_{\omega y}Y^{\mu} + C_{\omega z}Z^{\mu}, 
\eea
where the coefficient functions $C$ depend only on the proper time $\tau$. For convenience, we introduce a three-vector notation
\bea
\Ckv = (C_{kx},C_{ky},C_{kz}), \quad
\Cov = (C_{\omega x},C_{\omega y},C_{\omega z}),
\eea
and define the four-vector $t$ as the contraction of $t^{\mu \nu}$ with $k_{\nu}$,
\bea
t^{\mu} = t^{\mu \nu} k_{\nu} = \epsilon^{\mu \nu \alpha \beta} k_{\nu} U_{\alpha} \omega_{\beta}.
\eea
We note that $t^\mu$ is a second-order quantity, as it involves products of the components of $k$ and $\omega$. Combining the formulas given above, we can write
\begin{align}\begin{split}
k^{\mu} &= \LB C_{kz}  \sinh \eta, C_{kx}, C_{ky}, C_{kz} \cosh\eta \RB,  \\
\omega^{\mu} &= \LB C_{\omega z}  \sinh\eta, C_{\omega x}, C_{\omega y}, C_{\omega z} \cosh\eta \RB, \\
t^\mu &= \LB  V_z \sinh\eta, V_x, V_y,  V_z \cosh\eta \RB = V_x X^\mu + V_y Y^\mu + V_z Z^\mu,
\end{split}\end{align}
where we have again used the three-vector notation
\bea
\Vv = (V_x, V_y, V_z) = \Ckv \times \Cov .
\eea
For $\eta=0 \,\,\,(z=0)$, we obtain $\tv = \Vv$.

In hydrodynamic calculations, it is common to introduce the tensor $\Delta^{\mu \nu}$, which projects onto the space orthogonal to the flow vector, $\Delta^{\mu\nu} U_\mu =0$. Using the basis \EQn{eq:4_vectors}, we can write
\bea
\Delta^{\mu \nu} = g^{\mu \nu} - U^{\mu} U^{\nu}
= - X^\mu X^\nu - Y^\mu Y^\nu- Z^\mu Z^\nu.
\eea
Alongside this definition, we define the convective and transverse derivatives as follows
\bea
D = U^\mu \p_\mu, \qquad \nabla^\nu = \Delta^{\mu \nu} \p_\mu.  
\eea
One can easily verify that for any function of the proper time only, $f=f(\tau)$, the following holds: $Df(\tau) = \dd f(\tau)/d\tau$ and $\nabla^\mu f(\tau)=0$. We also note the relation
\bea
\p_\mu \Delta^{\mu \nu} =  - \f{U^\nu}{\tau}
\eea
and the trace $\Delta^\mu_{\,\,\,\mu}=3$.

\subsection{Differential identities}

Since we will be using the variables $\tau$ and $\eta$, we convert the derivatives according to the  following transformation rule~\CITn{Florkowski:2010zz}
\bea
\begin{bmatrix}
\partial_t&\\ 
\partial_x&\\ 
\partial_y&\\
\partial_z&
\end{bmatrix}
=
\begin{bmatrix}
\cosh\eta&\  0&\ \ 0&  -\f{1}{\tau} \sinh\eta& \\ 
0& \                 1&\ \ 0&  0& \\ 
0& \                 0&\ \ 1&  0&  \\
-\sinh\eta&\ 0&\ \ 0& \f{1}{\tau} \cosh\eta &
\end{bmatrix}
\begin{bmatrix}
\partial_{\tau}&\\ 
\partial_x&\\ 
\partial_y&\\
 \partial_{\eta}&
\end{bmatrix}.
\eea
Using the notation
\bea
{\dot f}(\tau) = \f{\dd f(\tau)}{\dd \tau}, \qquad 
 f^\prime(\eta) = \f{\dd f(\eta)}{\dd \eta},
\eea
where $f$ represents a differentiable function, we can derive several useful relations:
\begin{align}\begin{split}
U^\mu \p_\mu f(\tau) &= {\dot f}, \qquad  
U^\mu \p_\mu f(\eta) = 0, \\ 
k^\mu \p_\mu f(\tau) &= 0, \qquad 
k^\mu \p_\mu f(\eta) = \f{C_{kz}}{\tau} f^\prime(\eta),  \\ 
\omega^\mu \p_\mu f(\tau) &= 0, \qquad
\omega^\mu \p_\mu f(\eta) = \f{C_{\omega z}}{\tau} f^\prime(\eta),  \\ 
t^\mu \p_\mu f(\tau) &= 0, \qquad 
t^\mu \p_\mu f(\eta) = \f{V_z }{\tau}  f^\prime(\eta).
\end{split}\end{align}
Equations listed above are particularly useful for calculating gradients of thermodynamic quantities that are scalars and depend on $\tau$ only. Further identities required to derive hydrodynamic equations involve divergences of the vectors: $U^\mu, k^\mu, \omega^\mu$ and $t^\mu$, as well as the differential operators $U^\mu \p_\mu$, $k^\mu \p_\mu$, $\omega^\mu \p_\mu$, and $t^\mu \p_\mu$. We also need the directional derivatives $U^\mu \p_\mu A^\nu$, $k^\mu \p_\mu A^\nu$, $\omega^\mu \p_\mu A^\nu$, and $t^\mu \p_\mu A^\nu$, where $A^\nu$ is $U^\nu, k^\nu, \omega^\nu$, or $t^\nu$. All these operators and derivatives are provided explicitly in Appendix~\ref{sec:AppA}.

\section{Conservation laws for perfect fluid hydrodynamics}

\subsection{Baryon number conservation}

The baryon number conservation law has the form
\bel{eq:dN}
\p_\mu N^\mu (x)=0,
\eel
with the baryon number current defined by the formula~\CITn{Florkowski:2024bfw,Drogosz:2024gzv}
\bea
N^{\mu}= \bar{n}(T,\xi,k^2,\omega^2) U^{\mu} + n_t(T,\xi) t^{\mu}.
\eea
Explicit expressions for the functions $\bar{n}(T,\xi,k^2,\omega^2)$ and $n_t(T,\xi)$ are given in Appendix~\ref{sec:explicit}. Since the hydrodynamic variables $T, \xi, k^2$, and $\omega^2$ depend only on $\tau$, it can be readily verified using \EQ{eq:du_ud} that \EQ{eq:dN} leads to
\bea
 \f{\dd \bar{n}}{\dd \tau} +\f{\bar{n}}{\tau} = 0,
\eea
which has the following scaling solution
\bea
\bar{n} \LB T(\tau),\xi(\tau),k^2(\tau),\omega^2(\tau) \RB = \f{\tau_0}{\tau} \, \bar{n} \LB T(\tau_0),\xi(\tau_0),k^2(\tau_0),\omega^2(\tau_0) \RB,
\label{eq:scal}
\eea
where $\tau_0$ is the initial value of the proper time. 

\subsection{Energy and linear momentum conservation}

The energy-momentum conservation law is expressed by the formula
\bea
\partial_{\mu} T^{\mu \nu}=0,
\eea
where the energy-momentum tensor has the form~\CITn{Florkowski:2024bfw,Drogosz:2024gzv}
\bea
T^{\mu \nu} &=& \bar{\varepsilon} U^{\mu} U^{\nu} - \bar{P} \Delta^{\mu \nu} + P_k k^{\mu} k^{\nu} + P_{\omega} \omega^{\mu} \omega^{\nu} + P_t \left( t^{\mu} U^{\nu} + t^{\nu} U^{\mu} \right).
\eea
Explicit expressions for the functions appearing above are given in Appendix~\ref{sec:explicit}. Using the differential identities listed in the previous section and in Appendix~\ref{sec:AppA} we find
\begin{align}\begin{split}\label{eq:dTUXYZ}
\partial_{\mu} T^{\mu \nu} &= \LSB \dot{\bar{\varepsilon}} + \f{\bar{\varepsilon} + \bar{P}}{\tau} + \f{P}{\tau} \LB C^2_{kz} + C^2_{\omega z} \RB \RSB U^\nu \\
&+ \LSB \LB \dot{P_t} + \f{P_t}{\tau} \RB V_x + P_t \dot{V_x} \RSB X^\nu\\
&+ \LSB \LB \dot{P_t} + \f{P_t}{\tau} \RB V_y + P_t \dot{V_y} \RSB Y^\nu\\
&+ \LSB \LB \dot{P_t} + \f{P_t}{\tau} \RB V_z+
P_t \LB \dot{V_z} + \f{V_z}{\tau} \RB  
\RSB Z^\nu.
\end{split}\end{align}
Since the vectors $U, X, Y$, and $Z$ are linearly independent, all coefficients multiplying these vectors in the above equation  must individually vanish. It is useful to note that the terms involving $\Vv$ are quadratic corrections. Without such quadratic corrections the energy-momentum conservation is reduced to the first line in~\EQ{eq:dTUXYZ} only.

\subsection{Conservation of the spin part of the angular momentum}

We assume that the spin tensor fulfills the conservation law 
\bel{eq:DS_law}
\p_{\lambda} S^{\lambda, \mu \nu} = 0
\eel
and has the form~\CITn{Florkowski:2024bfw,Drogosz:2024gzv}
\bel{eq:S}
S^{\lambda, \mu \nu} &=& U^{\lambda} \left[ A \LB  k^\mu U^\nu - k^\nu U^\mu \RB + A_1 t^{\mu \nu} \right] \nn \\
&+& \f{A_3}{2} \left[ \LB t^{\lambda \mu} U^\nu - t^{\lambda \nu} U^\mu \RB + \LB \Delta^{\lambda \mu} k^\nu - \Delta^{\lambda \nu} k^\mu \RB \right],
\eel
where $A = A_3$ and $A_1$ are scalar functions depending only on $\tau$. Expanding \EQ{eq:DS_law} with the help of \EQ{eq:S}, we obtain the relation
\begin{align}\begin{split}\label{eq:dS}
\p_\lambda S^{\lambda, \mu \nu} &= \f{1}{\tau}
\left[ A \LB k^\mu U^\nu - k^\nu U^\mu \RB +A_1 t^{\mu \nu} \right] + ( A k^\mu 
\dot{)} U^\nu - ( A k^\nu \dot{)} U^\mu \\
&+ \dot{A}_1 t^{\mu \nu} + A_1 \dot{t}^{\mu \nu} 
+ \f{1}{2} \p_\lambda A_3 \LB t^{\lambda \mu} U^\nu - t^{\lambda \nu} U^\mu \RB  \\ 
&+ \f{A_3}{2} \Big[ \LB \p_\lambda t^{\lambda \mu} \RB U^\nu + t^{\lambda \mu} \p_\lambda U^\nu - \LB \p_\lambda t^{\lambda \nu} \RB U^\mu - t^{\lambda \nu} \p_\lambda U^\mu  \\
& - \f{U^\mu}{\tau} k^\nu + \Delta^{\lambda \mu} \p_\lambda k^\nu + \f{U^\nu}{\tau} k^\mu - \Delta^{\lambda \nu} \p_\lambda k^\mu \Big] = 0.
\end{split}\end{align}
To solve \EQ{eq:dS}, we contract it with each of the tensors: $X_\mu Y_\nu$,  $Y_\mu Z_\nu$,  $Z_\mu X_\nu$,  $U_\mu X_\nu$, $U_\mu Y_\nu$, and $U_\mu Z_\nu$ separately, using formulas from Appendix~\ref{sec:contract}. In this way, we obtain six equations, which can be written in matrix form as follows
\begin{equation}\label{eq:cs}
\begin{bmatrix}
A & 0 & 0 & 0 & 0 & 0 \\
0 & A & 0 & 0 & 0 & 0 \\
0 & 0 & A & 0 & 0 & 0 \\
0 & 0 & 0 & A_1 & 0 & 0 \\
0 & 0 & 0 & 0 & A_1  & 0 \\
0 & 0 & 0 & 0 & 0 & A_1
\end{bmatrix}
\begin{bmatrix}
\Dot{C}_{k x} \\
\Dot{C}_{k y} \\
\Dot{C}_{k z} \\
\Dot{C}_{\omega x} \\
\Dot{C}_{\omega y} \\
\Dot{C}_{\omega z} \end{bmatrix}=\begin{bmatrix}
Q_1 & 0 & 0 & 0 & 0 & 0 \\
0 & Q_1 & 0 & 0 & 0 & 0 \\
0 & 0 & Q_2 & 0 & 0 & 0 \\
0 & 0 & 0 & R_1 & 0 & 0 \\
0 & 0 & 0 & 0 & R_1  & 0 \\
0 & 0 & 0 & 0 & 0 & R_2
\end{bmatrix}
\begin{bmatrix}
{C}_{k x} \\
{C}_{k y} \\
{C}_{k z} \\
{C}_{\omega x} \\
{C}_{\omega y} \\
{C}_{\omega z} \end{bmatrix}, 
\end{equation}
where
\begin{align}\begin{split}
Q_1 &= -\left[\dot{A}+\frac{1}{\tau}
\left( A+ \frac{1}{2}A\right) \right], \qquad
Q_2 = -\left(\dot{A}+\frac{A}{\tau} \right),\\
R_1 &=- \left[\Dot{A_1}+\frac{1}{\tau}
\left(A_1 -\frac{1}{2} A \right) \right], \qquad
R_2 = -\left(\Dot{A_1} +\frac{A_1}{\tau}\right).
\label{eq:Q_R}
\end{split}\end{align}

\subsection{Classification of the solutions}

In the previous sections, we obtained eleven boost-invariant hydrodynamic equations for eight unknown functions of the proper time: $T(\tau), \xi(\tau), \Ckv(\tau)$ and $\Cov(\tau)$. The reason for this is an externally imposed boost-invariant form of the hydrodynamic flow that fixes three (usually independent) components of the four-vector $U$. We note that this problem did not appear in~\CIT{Florkowski:2019qdp}, where the second-order corrections to the baryon current and to the energy-momentum tensor were ignored. In that case, the energy-momentum conservation laws were reduced to a single equation, and we were dealing with eight equations for eight unknowns. The overdetermined system of equations obtained in this work can be reduced to a well defined one (where the number of equations is equal to the number of unknowns) if certain additional symmetry constraints are imposed. Below we discuss two special cases where the coefficients functions $\Ckv(\tau)$ and $\Cov(\tau)$ satisfy additional conditions.

\begin{figure}[t]
\subfloat[][\centering]{\includegraphics[width=0.4\textwidth]{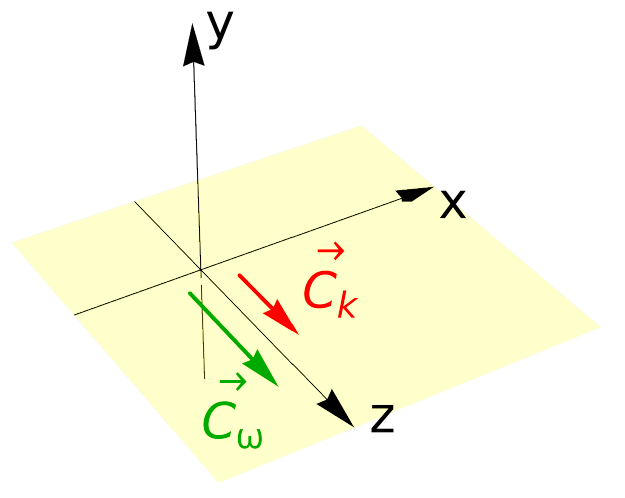} \label{fig:longitudinal}}
\qquad
\subfloat[][\centering]{\includegraphics[width=0.3\textwidth]{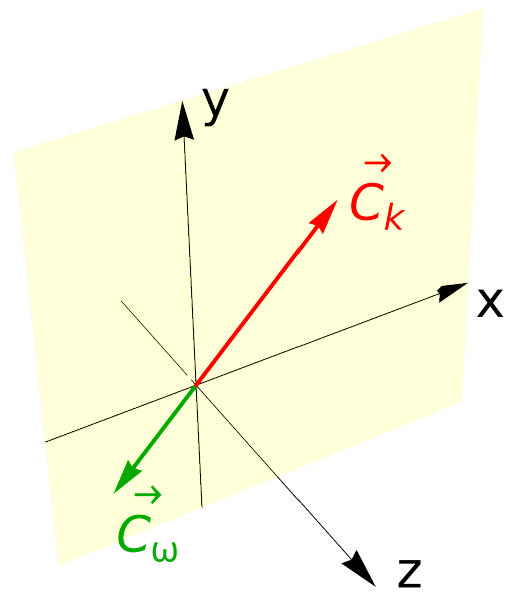} \label{fig:transverse}}
\caption{Schematic view of the longitudinal (a) and the transverse (b) configuration.}
\end{figure}

\subsubsection{Longitudinal configuration}

The structure of \EQ{eq:dTUXYZ} suggests that our system of eleven equations is reduced to eight equations in the case where the cross product $\Vv = \Ckv \times \Cov$ vanishes. The last condition can be satisfied for the spin configuration defined by the expressions
\bel{eq:L_cond}
\Ckv = (0,0, C_{kz}), \qquad \Cov = (0,0, C_{\omega z}).
\eel
See Fig.~\ref{fig:longitudinal} for its schematic representation.
In this case, we are in fact left with only four equations:
\bel{eq:LEQn}
\Dot{\bar{n}}  &=& - \f{\bar{n}}{\tau}, 
  \\
\dot{\bar{\varepsilon}} &=& - \f{\bar{\varepsilon} + \bar{P}}{\tau} - \f{P}{\tau} \LB C^2_{kz} + C^2_{\omega z} \RB, \label{eq:LEQe} \\
\Dot{C}_{k z} &=& \f{Q_2}{A} \, C_{k z}, \label{eq:LEQCk} \\
\Dot{C}_{\omega z} &=& \f{R_2}{A_1} \, C_{\omega z}.
\label{eq:LEQCo}
\eel
Using our definitions of the functions $Q_2$ and $R_2$, we can write formal solutions of the last two equations above as
\bel{eq:1_CkCoSOL}
C_{kz}(\tau) = \f{C_{kz}^0 A^0 \tau_0}{A \LB T(\tau),\xi(\tau) \RB \tau}, \qquad
C_{\omega z}(\tau) = \f{C_{\omega z}^0 A_1^0 \tau_0}{A_1 \LB T(\tau),\xi(\tau) \RB \tau},
\eel
where the index $0$ denotes the initial value of a given function. 

Substituting \EQSTWO{eq:LEQCk}{eq:LEQCo} into \EQSTWO{eq:LEQn}{eq:LEQe}, we obtain two coupled ordinary differential equations. This system of equations can be further simplified if we use~\EQ{eq:scal} instead of~\EQ{eq:LEQn}. In this case, we can reduce~\EQSM{eq:LEQn}{eq:LEQCo} to a single ordinary differential equation. Nevertheless, in practice we numerically solve~\EQSM{eq:LEQn}{eq:LEQCo} as a system of four equations for four unknowns: $T$, $\mu$, $C_{kz}$, and $C_{\omega z}$. The analytic results, such as \EQS{eq:1_CkCoSOL}, are of use as checks of numerical accuracy. 

\begin{figure}[t]
    \centering
    \includegraphics[width=1.\linewidth]{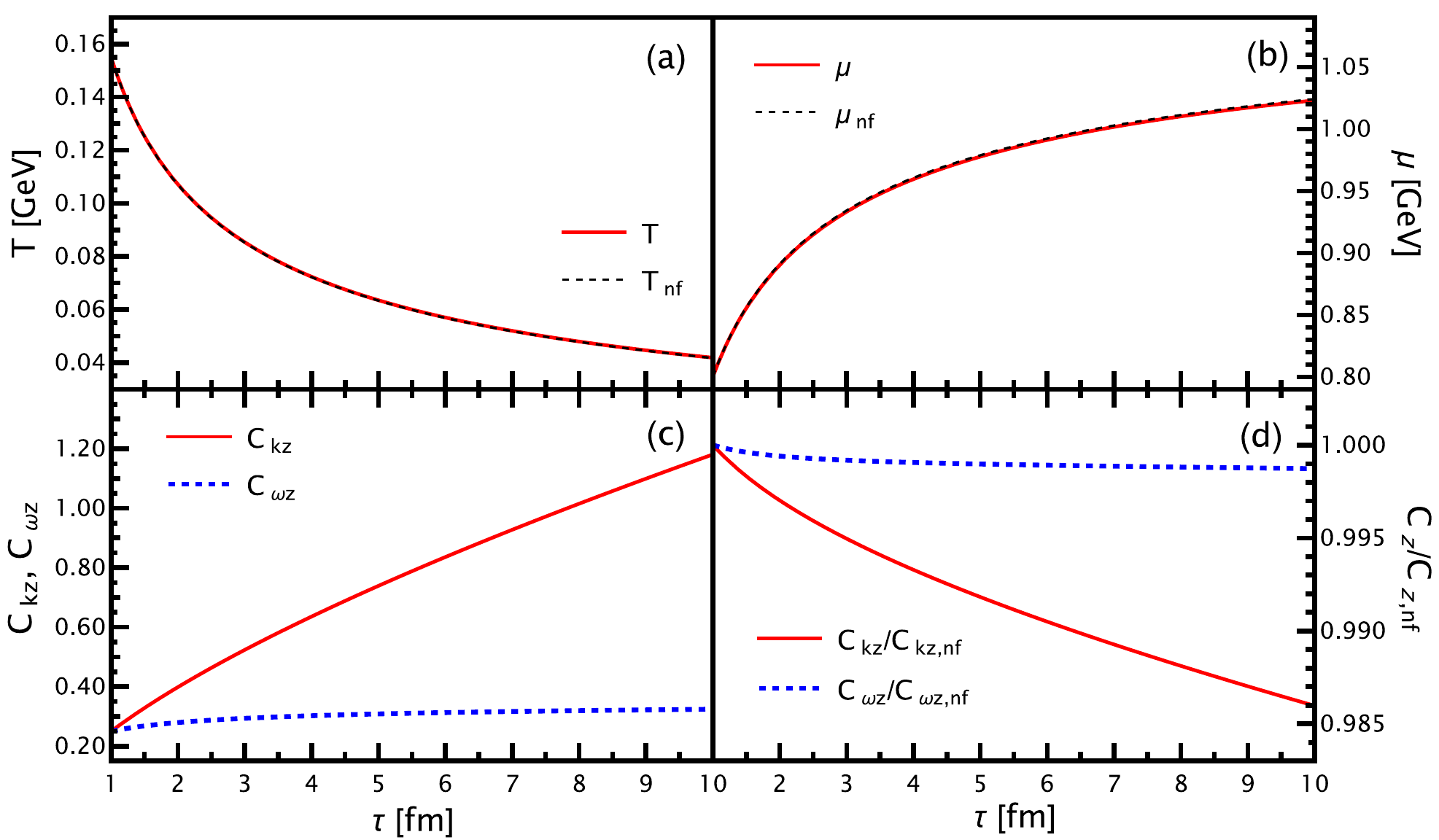}
    \caption{Longitudinal configuration with the initial values $C^0_{kz} = C^0_{\omega z} = 0.25$. Proper-time dependence of (a) temperature $T$, (b) baryon chemical potential $\mu$, (c) coefficients $C_{kz}$ and $C_{\omega z}$ and (d) their ratios to no-feedback results.}\label{fig:L_0.25}
\end{figure}

\begin{figure}[t]
    \centering
    \includegraphics[width=1.\linewidth]{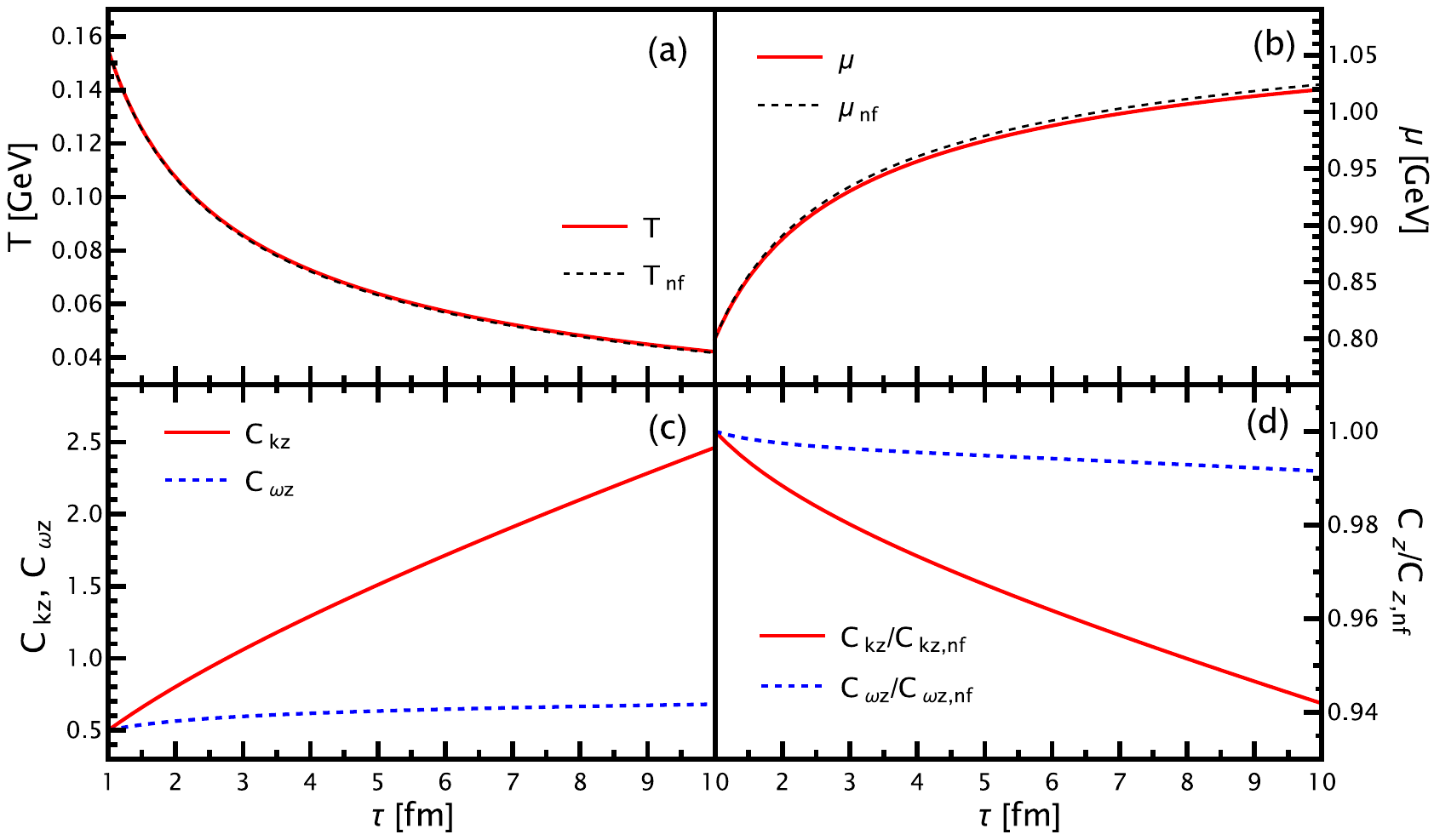}
    \caption{Same as Fig. \ref{fig:L_0.25} but with the initial values $C^0_{k z} = C^0_{\omega z} = 0.50$.}
    \label{fig:L_0.5}
\end{figure}

\begin{figure}[t]
    \centering
    \includegraphics[width=1.\linewidth]{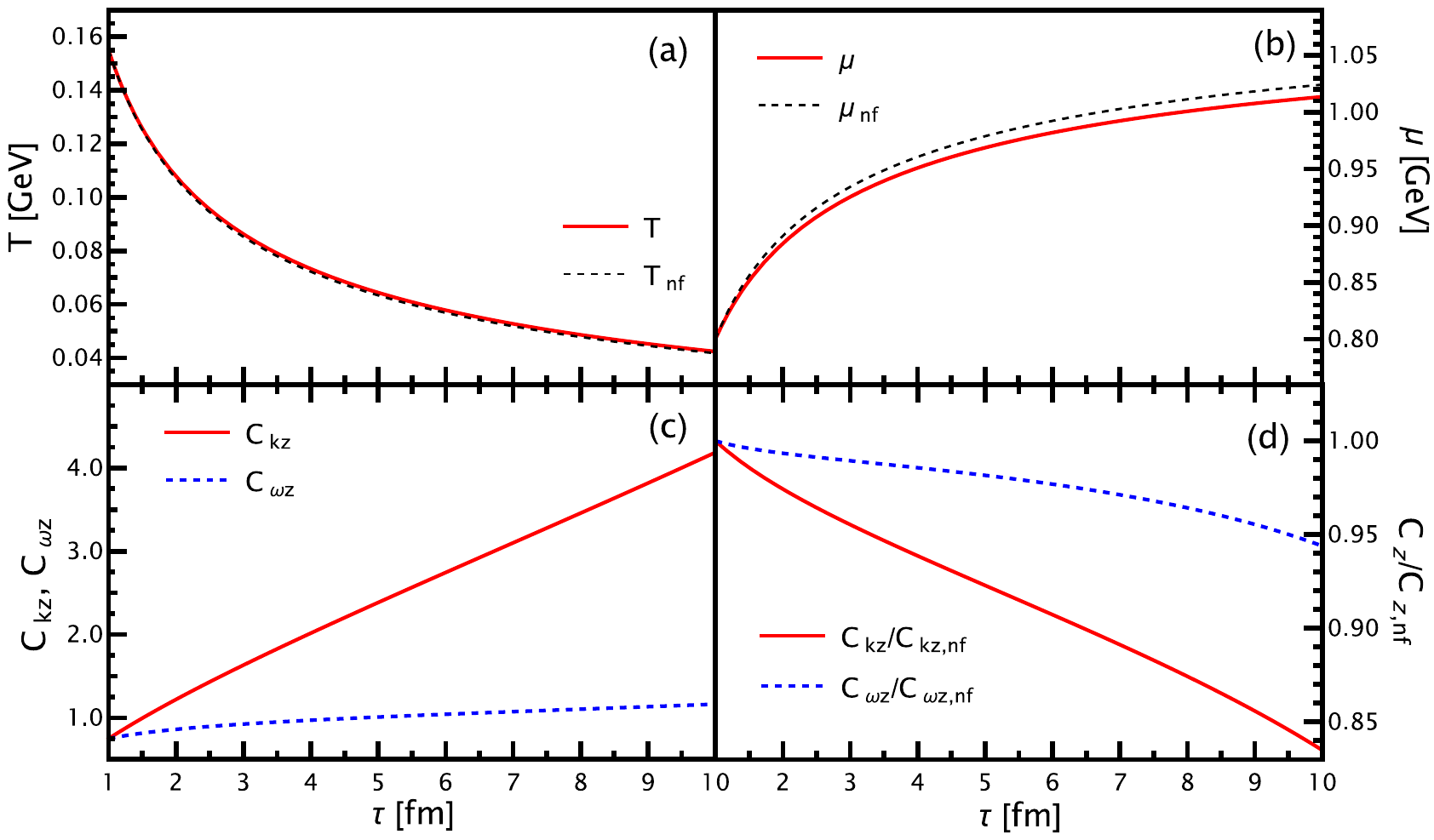}
    \caption{Same as Fig. \ref{fig:L_0.25} but with the initial values $C^0_{k z} = C^0_{\omega z} = 0.75$.}
    \label{fig:L_0.75}
\end{figure}

 
\begin{figure}[t]
    \centering
    \includegraphics[width=1.\linewidth]{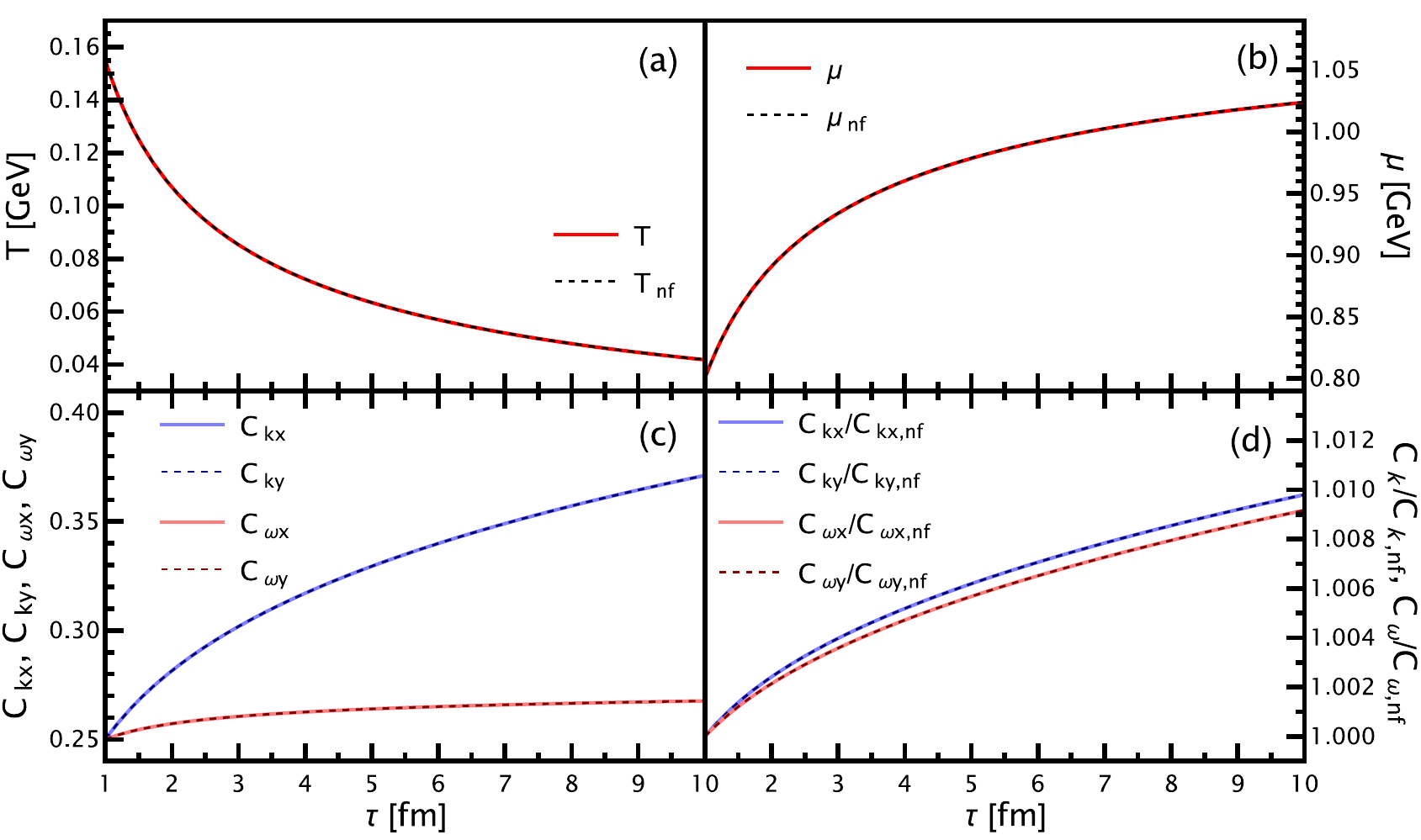}
    \caption{Transverse configuration with the initial values $C^0_{kx} = C^0_{ky} = C^0_{\omega x} = C^0_{\omega y} = 0.25$. Proper-time dependence of (a) temperature $T$, (b) baryon chemical potential $\mu$, (c) coefficients $C_{kx}$, $C_{ky}$, $C_{\omega x}$, and $C_{\omega y}$, and (d) their ratios to no-feedback results.}
    \label{fig:T_0.25}
\end{figure}

\begin{figure}[t]
    \centering
    \includegraphics[width=1.\linewidth]{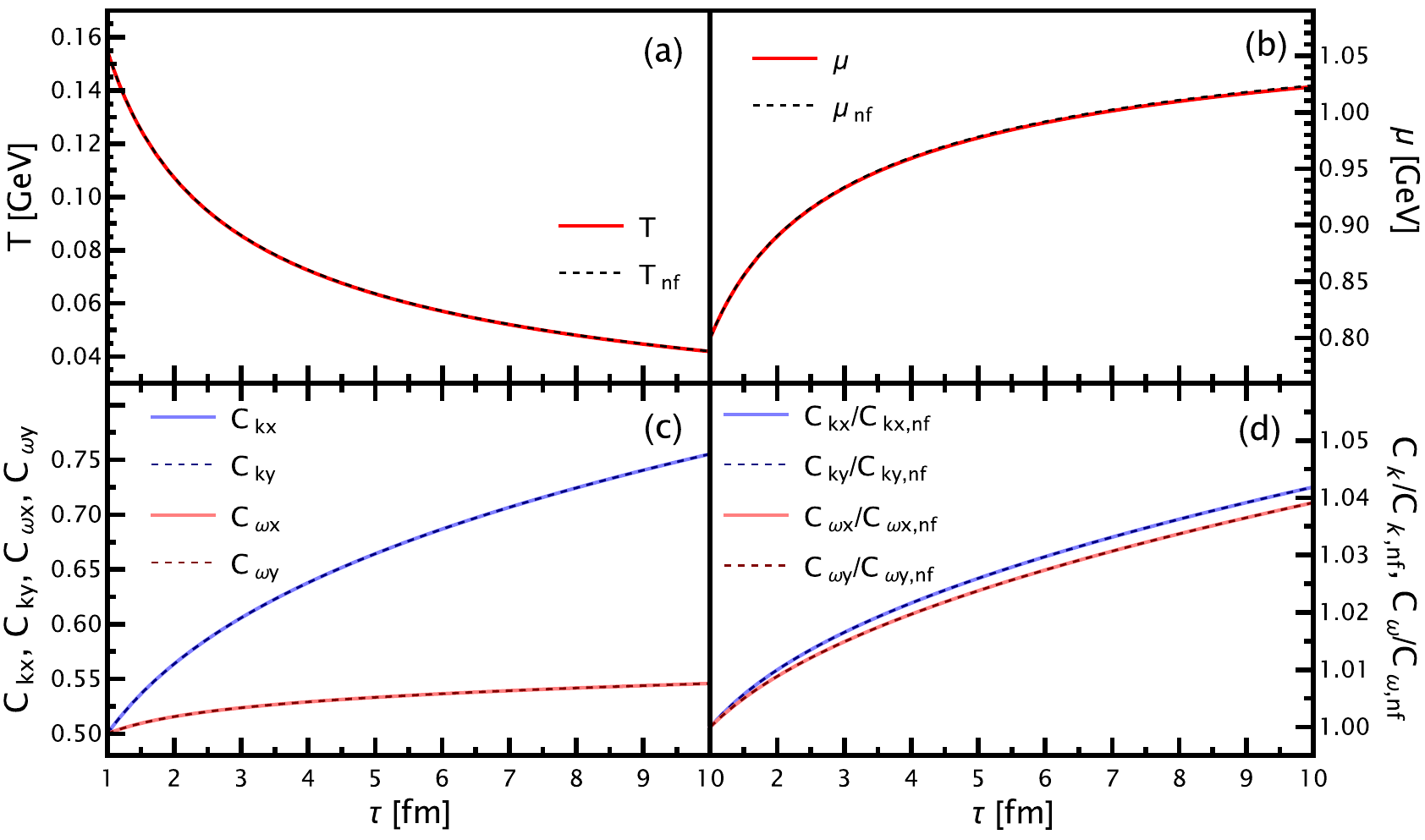}
    \caption{Same as Fig.~\ref{fig:T_0.25} but with the initial values $C^0_{kx} = C^0_{ky} = C^0_{\omega x} = C^0_{\omega y} = 0.50$.}
    \label{fig:T_0.5}
\end{figure}

\begin{figure}[t]
    \centering
    \includegraphics[width=1.\linewidth]{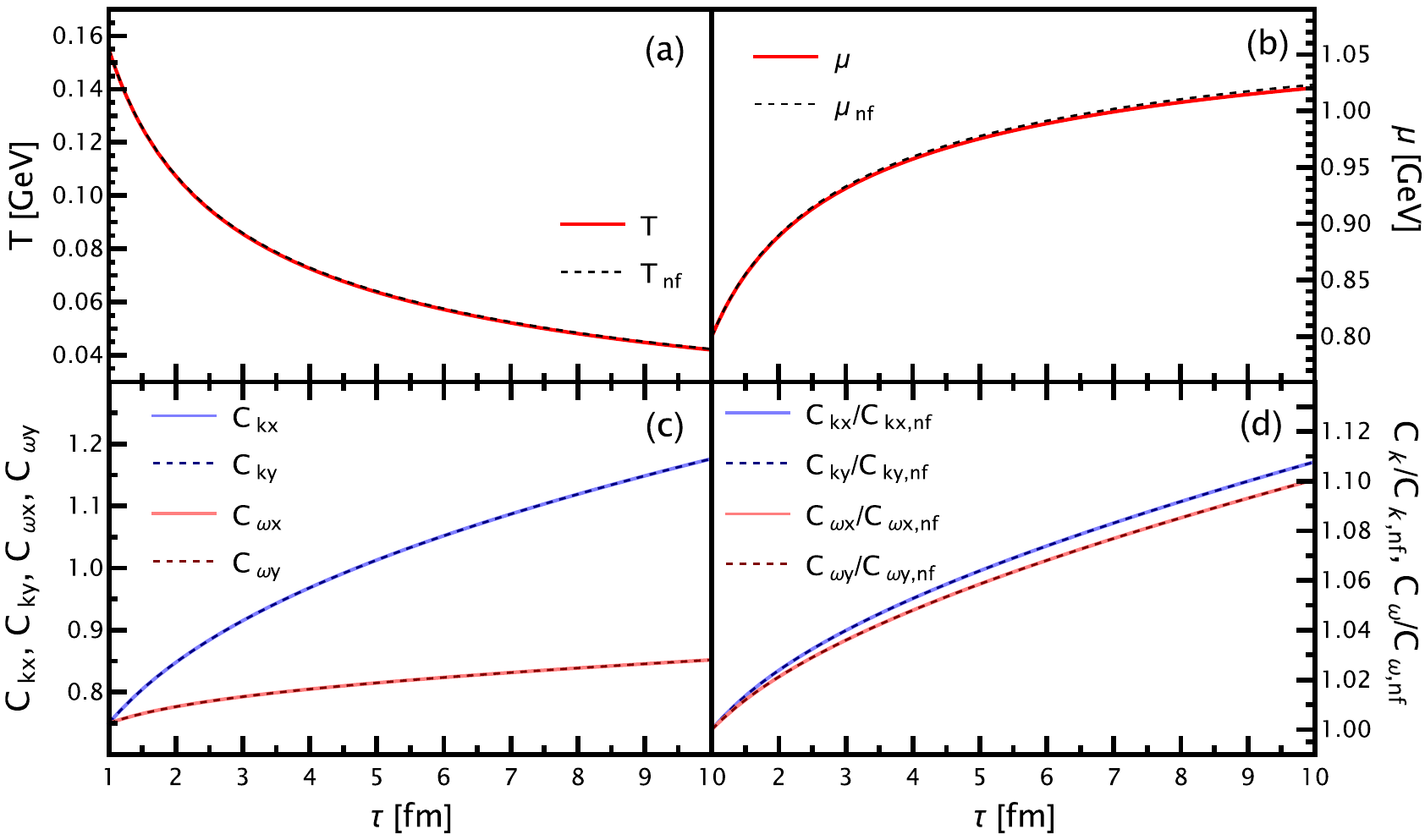}
    \caption{Same as Fig. \ref{fig:T_0.25} but with the initial values $C^0_{kx} = C^0_{ky} = C^0_{\omega x} = C^0_{\omega y} = 0.75$.}
    \label{fig:T_0.75}
\end{figure}
\subsubsection{Transverse configuration}
Another choice that also reduces the number of equations (with the condition $\Vv = 0$) is the ansatz
\bel{eq:T_postulate}
C_{kx} C_{\omega y} - C_{ky} C_{\omega x} = 0, \qquad C_{kz} = C_{\omega z} = 0,
\eel
which leads to the form~\footnote{One may note here that such a configuration was also considered in the context of purely longitudinal boost-invariant solutions of anomalous magnetohydrodynamics in Ref.~\CITn{Siddique:2019gqh}.}
\bel{eq:T_cond}
\Ckv = \LB C_{kx}, C_{ky}, 0 \RB, \qquad \Cov = \lambda \Ckv.
\eel
See Fig.~\ref{fig:transverse} for a schematic view of the configuration.

Using~\EQS{eq:cs}, one can check that~\EQS{eq:T_postulate} hold throughout the whole evolution of the system, provided they are satisfied at the initial time $\tau=\tau_0$. This is a straightforward consequence of the equation 
\bea
\f{d}{d \tau} \LB C_{k x} C_{\omega y} -  C_{k y} C_{\omega x} \RB = 0,
\eea
which directly follows from~\EQS{eq:cs}. In general, the proportionality parameter $\lambda$ in \EQ{eq:T_cond} is a function of the proper time $\tau$, $\lambda = \lambda(\tau)$. Below, as the initial condition at $\tau=\tau_0$, we use the value $\lambda (\tau_0) = 1$. 

\medskip
In the ``transverse'' case we are left with six differential equations:
\bel{eq:TEQne}
\Dot{\bar{n}}  &=& - \f{\bar{n}}{\tau},  \qquad \dot{\bar{\varepsilon}} = - \f{\bar{\varepsilon} + \bar{P}}{\tau},   \\
\Dot{C}_{k x} &=& \f{Q_1}{A} \, C_{k x}, \qquad \Dot{C}_{k y} = \f{Q_1}{A} \, C_{k y}, \label{eq:TEQCk} \\ 
\Dot{C}_{\omega x} &=& \f{R_1}{A_1} \, C_{\omega x}, \qquad
\Dot{C}_{\omega y} = \f{R_1}{A_1} \, C_{\omega y}. \label{eq:TEQCo} 
\eel
Owing to the rotational invariance in the transverse plane, the $x$ and $y$ components of the vector $\Ckv$ satisfy the same equations. Analogically to \EQS{eq:1_CkCoSOL}, we can write analytic solutions of \EQS{eq:TEQCk} as
\bel{eq:2_CkCoSOL}
C_{k x}= \f{C_{k x}^0 A^0 \tau_0}{A(T(\tau), \xi(\tau))  \tau^{3/2}}, \qquad C_{k y}= \f{C_{k y}^0 A^0 \tau_0}{A(T(\tau), \xi(\tau))  \tau^{3/2}}.
\eel
The $x$ and $y$ components of the vector $\Cov$ also satisfy the same equations, however, one cannot readily obtain an analytic solution in this case because the term $R_1/A_1$ cannot be written as a logarithmic derivative. We can write
\begin{equation}
\f{R_1}{A_1} = - \f{\dot{A}_1}{A_1} - \f{1}{\tau} \LB 1 - \f{1}{2} \f{A_3}{A_1} \RB = - \f{\dot{A}_1}{A_1} - \f{1}{\tau} \LB 1 + \f{K_3(z)}{z K_2(z) + 2K_3(z)} \RB \equiv - \f{\dot{A}_1}{A_1} - \f{1}{\tau} M(z),
\end{equation}
where the function $M(z)$ takes values from $1.0$ to $1.5$ and decreases with increasing $z$. It approaches unity for low temperatures ($z=m/T \gg 1$).


\section{Numerical results}
In this section we present the results of our numerical calculations for the longitudinal and transverse configurations, defined by~\EQSM{eq:LEQn}{eq:LEQCo} and~\EQSM{eq:TEQne}{eq:TEQCo}, respectively. We start the time evolution at the initial proper time $\tau_0 =$~1~fm and continue until the final time $\tau_f=10$ fm. Explicit expressions for the baryon current, energy-momentum tensor, and spin tensor are given in Appendix~\ref{sec:explicit}. The particle mass is taken to be that of the Lambda hyperon, $m=1.116$~GeV. The initial conditions for temperature and baryon chemical potential resemble those used in realistic simulations of heavy-ion collisions at moderate energies: $T_0=155$~MeV, $\mu_0=800$~MeV. The initial conditions for the coefficients $C$ are chosen to be 0.25, 0.50, and 0.75 (note that these are dimensional quantities in natural units adopted herein). 

In the case of the longitudinal configuration we deal only with the coefficients $C_{k z}$ and $C_{\omega z}$ and assume that their values are equal at the initial proper time. The numerical results for the initial values 0.25, 0.50, and 0.75 are shown in Figs.~\ref{fig:L_0.25}--\ref{fig:L_0.75}, respectively. Each of these figures consists of four panels: The upper left panel shows the proper-time dependence of temperature, the upper right one shows the time dependence of the chemical potential, the lower left one shows the time dependence of $C_{k z}$ and $C_{\omega z}$, while the lower right panel shows the time evolution of the ratios $C_{k z}/C_{k z, \hbox{\tiny nf}}$ and $C_{\omega z}/C_{\omega z, \hbox{\tiny nf}}$, where $C_{k z, \hbox{\tiny nf}}$ and $C_{\omega z, \hbox{\tiny nf}}$ are the results obtained in a previous analysis with no spin feedback included~\CITn{Florkowski:2019qdp}.

The results without the feedback are also shown in the upper panels of Figs.~\ref{fig:L_0.25}--\ref{fig:L_0.75} (dashed lines) for comparison. We observe that with growing magnitude of the initial values for $C_{k z}$ and $C_{\omega z}$ the discrepancies with the case without the feedback grow, which is an expected effect since the whole framework is based on the expansion in the coefficients $C_{k z}$ and $C_{\omega z}$. For the initial values $C^0_{k z} = C^0_{\omega z} = 0.75$, we find the differences of about 15\%. For initial values exceeding unity, the obtained results start to diverge with time, which reflects the overall breakdown of the expansion in the magnitude of the components of the spin polarization tensor.

Our results for the transverse configuration are shown in Figs.~\ref{fig:T_0.25}--\ref{fig:T_0.75}. In this case we consider four coefficients: $C_{kx}$, $C_{ky}$, $C_{\omega x}$, and $C_{\omega y}$. Their initial values at $\tau=\tau_0$ are assumed to be the same and, similarly to the longitudinal configuration, taken to be 0.25, 0.50, and 0.75. In the lower left panels of Figs.~\ref{fig:T_0.25}--\ref{fig:T_0.75} we observe that the $\omega$-coefficients behave differently from the $k$-coefficients. This difference is a direct consequence of different forms of \EQSTWO{eq:TEQCk}{eq:TEQCo}.

Similarly to the longitudinal configuration, we find that deviations from the case without the spin feedback grow with increasing magnitude of the initial conditions assumed for the coefficients $C$. Interestingly, in the case of the longitudinal configuration, the feedback results in the decrease of the coefficients, while in the case of the transverse configuration, it leads to their increase. In this case also we find divergencies to occur for initial values exceeding unity. 

\medskip
\noindent
\section{Summary and conclusions}

In this work, we have studied for the first time an extension of perfect spin hydrodynamics that includes second-order corrections in the spin polarization tensor. Such corrections lead to feedback of the spin degrees of freedom on the hydrodynamic evolution. To check possible effects of such corrections, we have considered for simplicity a one-dimensional boost-invariant expansion. 

We have found that the presence of the second-order corrections restricts the allowed forms of the spin polarization tensor, leading to the acceptable ``longitudinal'' and ``transverse'' configurations (defined in the main text). However, as long as the magnitude of the spin polarization tensor remains small, the allowed spin dynamics differs very little from that found previously in the case without the second-order corrections (spin feedback)~\CITn{Florkowski:2019qdp}. For the initial values of the spin tensor components that exceed unity, we observe a breakdown of our scheme, as expected since the whole formalism relies on the expansion up to the second-order terms that should be small. 

In the future, it would be interesting to use the boost-invariant geometry to study the effects of dissipation as defined, for example, in Refs.~\CITn{Florkowski:2024bfw,Drogosz:2024gzv}—in particular, to trace the role of asymmetric components of the energy-momentum tensor that may appear out of equilibrium and are responsible for spin-orbit interactions.

\bigskip
\noindent
{\it Acknowledgements.} The authors thank Samapan Bhadury, Mykhailo Hontarenko and Valeriya Mykhaylova for useful comments and discussions. This work was supported in part by the Polish National Science Centre (NCN) Grants No. 2022/47/B/ST2/01372 and No. 2018/30/E/ST2/00432.

%

\appendix

\section{Differential identities used to derive hydrodynamic equations}
\label{sec:AppA}

We obtain the following divergences and directional derivatives:
\begin{align}\begin{split}\label{eq:du_ud}
\p_\mu U^\mu &= \f{1}{\tau},\qquad U^\mu \p_\mu = \p_\tau,  \\
\p_\mu k^\mu &= 0, \qquad k^\mu \p_\mu = \f{C_{kz}}{\tau} \p_\eta, \\ 
\p_\mu \omega^\mu &= 0, \qquad \omega^\mu \p_\mu = \f{C_{\omega z}}{\tau} \p_\eta, \\
\p_\mu t^\mu &= 0, \qquad t^\mu \p_\mu =\f{V_z}{\tau} \p_\eta.
\end{split}\end{align}
In the derivation of hydrodynamic equations, it is also very convenient to use the derivatives of the hydrodynamic flow $U$,
\begin{align}\begin{split}
U^\mu \p_\mu U^{\nu} &= a^\nu = 0, \qquad 
k^\mu \p_\mu U^{\nu} = \f{C_{kz}}{\tau} Z^\nu, \\
\omega^\mu \p_\mu U^{\nu} &= \f{C_{\omega z}}{\tau} Z^\nu, \qquad
t^\mu \p_\mu U^{\nu} = \f{V_z}{\tau} \, Z^\nu,
\end{split}\end{align}
the derivatives of the four-vector $k$,
\begin{align}\begin{split}
U^\mu \p_\mu k^{\nu} = {\dot C_{kx}} X^\nu + {\dot C_{ky}} Y^\nu + {\dot C_{kz}} Z^\nu, \qquad
k^\mu \p_\mu k^{\nu} = \f{C_{kz}^2}{\tau} U^\nu, \\
\omega^\mu \p_\mu k^\nu = \f{C_{k z} C_{\omega z}}{\tau} U^\nu, \qquad
t^\mu \p_\mu k^\nu = \f{C_{kz} V_z}{\tau} \, U^\nu, 
\end{split}\end{align}
the derivatives of the four-vector $\omega$,
\begin{align}\begin{split}
U^\mu \p_\mu \omega^{\nu} = {\dot C}_{\omega x} X^\nu + {\dot C}_{\omega y} Y
^\nu + {\dot C}_{\omega z} Z^\nu, \qquad  
k^\mu \p_\mu \omega^{\nu} = \f{C_{k z} C_{\omega z}}{\tau} U^\nu, \\ 
\omega^\mu \p_\mu \omega^{\nu} = \f{C_{\omega z}^2}{\tau} U^\nu, \qquad
t^\mu \p_\mu \omega^\nu = \f{C_{\omega z} V_z}{\tau} U^\nu, 
\end{split}\end{align}
and the derivatives of the four-vector $t$,
\begin{align}\begin{split}
U^\mu \p_\mu t^\nu = {\dot V}_x X^\nu + {\dot V}_y Y^\nu + {\dot V}_z Z^\nu, \qquad
k^\mu \p_\mu t^\nu =  \f{C_{k z}  V_z}{\tau} \, U^\nu, \\  
\omega^\mu \p_\mu t^\nu =  \f{C_{\omega z}  V_z}{\tau} \, U^\nu, \qquad
t^\mu \p_\mu t^\nu =  \f{ V_z^2}{\tau} \, U^\nu.
\end{split}\end{align}

\section{Baryon current, energy-momentum tensor, and spin tensor decomposition \label{sec:explicit}}

The coefficients in the baryon current decomposition have the form
\bea
\bar{n} = n_0 +n_{2}^k + n_2^{\omega},
\eea
where
\begin{align}\begin{split}
n_0 &= \frac{2 \sinh \xi}{\pi^2}z^2 T^3 K_2(z), \\
n_2^k &= -\frac{2 \spin^2 \sinh \xi}{3 \pi^2}z T^3 K_3(z) k^2, \\
n_2^{\omega} &= -\frac{ \spin^2 \sinh \xi}{3 \pi^2}z T^3 (z K_2(z) + 2K_3(z))\omega^2, 
\end{split}\end{align}
and
\bel{eq:nt}
n_t &=& -\frac{2 \spin^2 \sinh \xi}{3 \pi^2}z T^3 K_3(z).
\eel
Here $n_0$ describes the baryon density of spinless relativistic gas, $\xi=\mu/T$ is the ratio of the baryon chemical potential and temperature, $z=m/T$, and $K_n$ denotes the modified Bessel function of the second type.

The coefficients in the energy-momentum tensor decomposition are given by~\CITn{Florkowski:2024bfw,Drogosz:2024gzv}
\begin{align}\begin{split}
\bar{\varepsilon} &= \varepsilon_0 + \varepsilon_2^k + \varepsilon_2^{\omega}, \\
\bar{P} &= P_0 +P_2^k +P_2^{\omega}, \\
P_k &= P_{\omega} \equiv P,
\end{split}\end{align}
where $\varepsilon_0$ and $P_0$ are the energy density and pressure of a relativistic gas of spinless particles, respectively, and
\begin{align}\begin{split}
\varepsilon_0 &= \frac{2 \hbox{cosh}(\xi)}{\pi^2} z^2 T^4 \left[ z K_3(z) - K_2(z) \right], \\
\varepsilon_2^k &= -\frac{2 \spin^2 \hbox{cosh}(\xi)}{3 \pi^2} z T^4 \left[ z K_2(z) + 5K_3(z) \right] k^2, \\
\varepsilon_2^{\omega} &= -\frac{ \spin^2 \hbox{cosh}(\xi)}{3 \pi^2} z T^4 \left[ z K_2(z) + (z^2 + 10) K_3(z) \right] \omega^2, 
\end{split}\end{align}

\begin{align}\begin{split}
P_0 &= \frac{2 \hbox{cosh}(\xi)}{\pi^2} z^2 T^4 K_2(z), \\
P_2^k &= -\frac{4 \spin^2 \hbox{cosh}(\xi)}{3 \pi^2} z T^4 K_3(z) k^2, \\
P_2^{\omega} &= -\frac{ \spin^2 \hbox{cosh}(\xi)}{3 \pi^2} z T^4 \left[ z K_2(z) + 4 K_3(z) \right] \omega^2, \\
P_t &= \frac{4 \spin^2 \hbox{cosh}(\xi)}{3 \pi^2} z T^4 \left[K_3(z) - z K_4(z) \right], \\
P_k &= P_{\omega} = P =  -\frac{2 \spin^2 \hbox{cosh}(\xi)}{3 \pi^2} z T^4 K_3(z).
\end{split}\end{align}

The coefficients $A$, $A_1$, $A_2$ and $A_3$ that appear in the spin tensor~\EQn{eq:S} are
\begin{align}\begin{split}
A &= A_1 - \f{A_2}{2} - A_3 = A_3, \\
A_1 &= \f{2\spin^2 \cosh \xi}{3 \pi^2} z T^3 \left[ z K_2(z) + 2 K_3(z) \right],\\
A_2 &= \f{4 \spin^2 \cosh \xi}{3 \pi^2}z^2 T^3 K_4(z), \\
A_3 &= -\f{4 \spin^2 \cosh \xi}{3 \pi^2} z T^3 K_3(z).
\end{split}\end{align}

\section{Conservation of the spin tensor and contractions with four-vectors} \label{sec:contract}

The following identities are useful when contracting~\EQ{eq:dS} with $X_\mu Y_\nu$, $Y_\mu Z_\nu$ and $Z_\mu X_\nu$
\begin{align}\begin{split}
X_\mu Y_\nu t^{\mu \nu} &= -C_{\omega z}, \qquad Y_\mu Z_\nu t^{\mu \nu} = -C_{\omega x} \qquad Z_\mu X_\nu t^{\mu \nu} = -C_{\omega y}, \\
X_\mu Y_\nu \dot{t}^{\mu \nu} &= -\dot{C}_{\omega z}, \qquad Y_\mu Z_\nu \dot{t}^{\mu \nu} = -\dot{C}_{\omega x} \qquad Z_\mu X_\nu \dot{t}^{\mu \nu} = -\dot{C}_{\omega y},
\end{split}\end{align}
\begin{align}\begin{split}
X_\mu Y_\nu t^{\lambda \mu} \p_\lambda U^\nu &= 0, \qquad
Y_\mu Z_\nu t^{\lambda \mu} \p_\lambda U^\nu = \f{C_{\omega x}}{\tau}, \qquad
Z_\mu X_\nu t^{\lambda \mu} \p_\lambda U^\nu = 0, \\
X_\mu Y_\nu t^{\lambda \nu} \p_\lambda U^\mu &= 0, \qquad 
Y_\mu Z_\nu t^{\lambda \nu} \p_\lambda U^\mu = 0, \qquad
Z_\mu X_\nu t^{\lambda \nu} \p_\lambda U^\mu = - \f{C_{\omega y}}{\tau},
\end{split}\end{align}
\begin{align}\begin{split}
X_\mu Y_\nu \Delta^{\lambda \mu} \p_\lambda k^\nu &= 0, \qquad
Y_\mu Z_\nu \Delta^{\lambda \mu} \p_\lambda k^\nu = 0, \qquad
Z_\mu X_\nu \Delta^{\lambda \mu} \p_\lambda k^\nu = 0, \\
X_\mu Y_\nu \Delta^{\lambda \nu} \p_\lambda k^\mu &= 0, \qquad 
Y_\mu Z_\nu \Delta^{\lambda \nu} \p_\lambda k^\mu = 0, \qquad
Z_\mu X_\nu \Delta^{\lambda \nu} \p_\lambda k^\mu = 0.
\end{split}\end{align}

The following formulas are useful when multiplying~\EQ{eq:dS} by the tensors $U_\mu X_\nu$, $U_\mu Y_\nu$ and $U_\mu Z_\nu$
\begin{align}\begin{split}
U_\mu X_\nu t^{\mu \nu} &= 0, \qquad U_\mu Y_\nu t^{\mu \nu} = 0, \qquad U_\mu Z_\nu t^{\mu \nu} = 0, \\
U_\mu X_\nu \dot{t}^{\mu \nu} &= 0, \qquad U_\mu Y_\nu \dot{t}^{\mu \nu} = 0, \qquad U_\mu Z_\nu \dot{t}^{\mu \nu} = 0,
\end{split}\end{align}
\begin{align}\begin{split}
U_\mu X_\nu t^{\lambda \mu} \p_\lambda U^\nu &= 0, \qquad
U_\mu Y_\nu t^{\lambda \mu} \p_\lambda U^\nu = 0, \qquad
U_\mu Z_\nu t^{\lambda \mu} \p_\lambda U^\nu = 0, \\
U_\mu X_\nu t^{\lambda \nu} \p_\lambda U^\mu &= 0, \qquad 
U_\mu Y_\nu t^{\lambda \nu} \p_\lambda U^\mu = 0, \qquad
U_\mu Z_\nu t^{\lambda \nu} \p_\lambda U^\mu = 0,
\end{split}\end{align}
\begin{align}\begin{split}
U_\mu X_\nu\Delta^{\lambda \mu} \p_\lambda k^\nu &= 0, \qquad
U_\mu Y_\nu \Delta^{\lambda \mu} \p_\lambda k^\nu = 0, \qquad
U_\mu Z_\nu \Delta^{\lambda \mu} \p_\lambda k^\nu = 0, \\
U_\mu X_\nu \Delta^{\lambda \nu} \p_\lambda k^\mu &= 0, \qquad 
U_\mu Y_\nu \Delta^{\lambda \nu} \p_\lambda k^\mu = 0, \qquad
U_\mu Z_\nu \Delta^{\lambda \nu} \p_\lambda k^\mu = \f{C_{k z}}{\tau},
\end{split}\end{align}
where the orthogonality relations $U_\mu t^{\mu \nu} = U_\nu t^{\mu \nu} = 0$ and $U_\mu \dot{t}^{\mu \nu} = U_\mu \varepsilon^{\mu \nu \gamma \delta} U_\gamma \dot{\omega}_\delta = 0$ hold.

\end{document}